\documentstyle[12pt]{article}
\newcommand{\newsubsection}[1]{
\vspace{1cm}
\pagebreak[3]
\addtocounter{subsection}{1}
\addcontentsline{toc}{subsection}{\protect
\numberline{\arabic{section}.\arabic{subsection}}{#1}}
\noindent{\arabic{subsection}. \sc  #1}                 
\nopagebreak
\vspace{2mm}
\nopagebreak}
\renewcommand{\theequation}{%\thesection.
\arabic{equation}}

\newlength{\extraspace}
\newlength{\extraspaces}
\newcommand{\be}{\begin{equation}
\addtolength{\abovedisplayskip}{\extraspaces}
\addtolength{\belowdisplayskip}{\extraspaces}
\addtolength{\abovedisplayshortskip}{\extraspace}
\addtolength{\belowdisplayshortskip}{\extraspace}}
\newcommand{\ee}{\end{equation}}
\newcommand{\ba}{\begin{eqnarray}
\addtolength{\abovedisplayskip}{\extraspaces}
\addtolength{\belowdisplayskip}{\extraspaces}
\addtolength{\abovedisplayshortskip}{\extraspace}
\addtolength{\belowdisplayshortskip}{\extraspace}}
\newcommand{\ea}{\end{eqnarray}}
\newcommand{\nonu}{\nonumber \\[2mm]}
\newcommand{\is}{&  \! =  \! &}
\newcommand{\twomatrix}[4]{{\left(\begin{array}{cc}#1 & #2\\
#3 & #4 \end{array}\right)}}

\newcommand{\ie}{{\it i.e.\ }}

\newcommand{\Tr}{{\rm Tr}}

\newcommand{\half}{{\textstyle{1\over 2}}}

\newcommand{\Z}{{\bf Z}}

\newcommand{\ra}{\rightarrow}

\newcommand{\Lbar}{{\overline{L}}}

\def\a{\alpha} 
\def\b{\beta}

\def\adot{{\dot\a}}
\def\bdot{{\dot\b}}
\def\th{\theta}
\newcommand{\R}{{\bf R}}
\newcommand{\xX}{\mbox{\it x}}
\newcommand{\sS}{{\lambda}}
\newcommand{\sSbar}{{\overline\lambda}}
\renewcommand{\d}{{\partial}}

\newcommand{\id}{{\bf 1}}

\newcommand{\ttr}{\! {\rm tr}\, }
\newcommand{\tr}{{\rm tr}}

\input epsf
\newcounter{fignum}
\newcommand{\figuurnum}{\arabic{fignum}}

\newcommand{\figuurplus}[3]{
\addtocounter{fignum}{1}
\addcontentsline{lof}{figure}{\protect
\numberline{\arabic{section}.\arabic{fignum}}{#3}}
\hspace{-3mm}{\it fig.}\ \figuurnum.
\begin{figure}[t]\begin{center}
\leavevmode\hbox{\epsfxsize=#2 \epsffile{#1.eps}}\\[5mm]
\parbox{10cm}{\small \bf Fig.\ \figuurnum: \it #3}
\end{center} \end{figure}\hspace{-1.5mm}}

\newcommand{\LL}{_{\! {}_{L}}}

\newcommand{\RR}{_{\! {}_{R}}}

\begin{document}

\thispagestyle{empty}

\begin{flushright}
{\sc March 1997}\\
{\sc cern-th/97-49}\\
%{\sc thu-97/06}\\
{\sc utfa-97/07}\\
\end{flushright}

\begin{center}
{\Large\sc{5D Black Holes and Matrix Strings}}\\[13mm]

{\sc Robbert Dijkgraaf}\\[2.5mm]
{\it Department of Mathematics}\\
{\it University of Amsterdam, 1018 TV Amsterdam}\\[4mm]

{\sc Erik Verlinde}\\[2.5mm]
{\it TH-Division, CERN, CH-1211 Geneva 23}\\[1mm]
and\\[.1mm]
{\it Institute for Theoretical Physics}\\
{\it Universtity of Utrecht, 3508 TA Utrecht}\\[3mm]

and\\[2mm]

{\sc  Herman Verlinde}\\[2.5mm]
{\it Institute for Theoretical Physics}\\
{\it University of Amsterdam, 1018 XE Amsterdam} \\[1.8cm]

{\sc Abstract}

\end{center}

\noindent
We derive the world-volume theory, the (non)-extremal entropy and
background geometry of black holes and black strings constructed out
of the NS IIA fivebrane within the framework of matrix theory. The CFT
description of strings propagating in the black hole geometry arises
as an effective field theory.

\noindent

\vfill

\newpage

\newsubsection{Introduction}

According to the matrix theory proposal of \cite{banks}
non-perturbative type II string dynamics can be captured by means of
an appropriate large $N$ limit of $U(N)$ supersymmetric quantum
mechanics. This proposal originated in the description of D-particles
and their bound states \cite{polch,bound,ulf}.  The relation with the
fundamental string degrees of freedom appears most manifestly in the
representation of matrix theory as $1+1$ dimensional supersymmetric
Yang-Mills theory, described by the Hamiltonian
\ba
H \is \oint \! d\sigma\, \tr\Bigl(\Pi_i^2 +
DX_i^2 +\theta^T \gamma^9 D \theta\nonumber \\[3mm] & & \qquad +
{1\over g_s} \theta^T\gamma^i[X_i,\theta] + {1 \over g_s^2} \,(E^2 +
[X^i,X^j]^2)\Bigr)  
\label{H}
\ea 
defined on the circle $0 \leq \sigma < 2\pi$.
Here $\Pi_i$ denotes the conjugate field to $X^i$, 
$E$ is the electric field and $D$ is the covariant derivative in 
the $\sigma$-direction. As shown in
\cite{banks}, \cite{motl}-\cite{matrix-string}, by promoting the string
fields $X^i,\th^\a,\th^\adot$ to hermitian $N\times N$
matrices one effectively achieves a second quantized description
of the free type IIA string. Furthermore, in the weak coupling 
limit\footnote{Here
we in fact mean the  strong coupling limit of the SYM model, 
which was shown in \cite{motl,banks-seiberg,matrix-string} to 
correspond to the limit of weakly interacting strings.} one reproduces 
the conventional perturbative interactions. This relation with
perturbative string theory strengthens the belief that the matrix string
possesses a ten-dimensional Lorentz invariance.

The matrix string formalism quite naturally combines the perturbative
string degrees of freedom with non-perturbative excitations such as
D-branes and NS fivebranes. The latter configurations can arise
because after toroidal compactifications matrix theory becomes
equivalent to higher dimensional supersymmetric large $N$ gauge theory
\cite{banks,herman,wati}. This in particular opens up the possibility
of adding new charged objects, essentially by considering gauge field
configurations that carry non-trivial fluxes or other topological
quantum numbers in the extra compactified directions.  It is not yet
clear, however, whether the matrix language can give an exhaustive
description of all non-perturbative excitations in string theory.  

To investigate this question, we will in this paper consider the
compactification to 4+1 and 5+1 dimensions. For the 4+1 dimensional case,
the non-perturbative spectrum of string theory was first analyzed by
Strominger and Vafa \cite{vafa-strominger}. They computed the
degeneracy of point-like BPS states as a function of their charge and
found a striking confirmation of the Bekenstein-Hawking entropy
formula \cite{bh} for extremal black holes. In subsequent works
\cite{blackhole2} a beautiful and simple effective description was
developed for the charged black hole degrees of freedom in terms of a
gas of relativistic strings. Our aim here
is to make contact with these results from the point of view of matrix
string theory. Closely related ideas have been put forward in the
recent paper \cite{li-martinec}.

In the original work of \cite{vafa-strominger} the effective strings
arose from the D-string confined to a D 5-brane.  Here we will
consider the dual BPS configuration with only NS charges made from
bound states of NS fivebranes with fundamental strings. Apart from
this, the formalism will in fact be rather similar to that of
\cite{vafa-strominger}. We will see that matrix theory gives a natural
unified description of fundamental strings, D-branes and solitonic
fivebranes. The individual objects correspond to submatrices that
encode the internal degrees of freedom. For the fivebrane these
degrees of freedom indeed combine into a second-quantized effective
string theory on the world-volume. 

The interactions between the different objects in matrix theory are
mediated by off-diagonal components of the matrices, which are the
analogues of the open strings stretching between different D-branes.
We will use this fact to probe the black string geometry by
considering its effect on the propagation of a type IIA matrix
string. This calculation is analogous to the D-string probe analysis
of five-dimensional black holes considered in \cite{dstring-probe} and
is related to the work of \cite{berkooz-douglas}. We will show how
the familiar CFT description of the NS fivebrane appears as an effective
description.

\newsubsection{Fivebranes and black strings}

In matrix string theory, longitudinal NS fivebranes that extend in the 
light-cone directions
%\footnote{This is one of the
%serious short comings of matrix theory in its present formulation.
%Hopefully it is just an artifact of the light-cone formulation, and
%not an indication of a break-down of ten-dimensional Lorentz
%invariance.}. 
correspond to configurations with a non-vanishing topological quantum number
\cite{ganor,branes}
\be
\label{wplus}
W^+_{ijkl} = \oint \! d\sigma\; \ttr X^{[i}X^jX^kX^{l]}. 
\ee 
This charge can naturally become non-vanishing when we compactify four of the 
transversal coordinates, so that $W^+_{ijkl}$ measures the winding 
number $k$ of the fivebrane around the corresponding four-torus.
In the matrix string theory Hamiltonian (\ref{H}), such a
compactification is described via the identification of the compact
$X^i$ as the covariant derivatives in a $5+1$ dimensional super
Yang-Mills theory on $S^1\times T^4$. 
In gauge theory language, the longitudinal
fivebrane appears as a YM instanton configuration, and the wrapping
number $k$ translates into the instanton charge on the $T^4$.  Note
that in $5+1$ dimensions these instanton solutions represent
string-like objects.

In the remaining six uncompactified dimensions, the wrapped 
fivebrane looks like an 
infinitely long string. This becomes a black string with a finite horizon 
area per unit length, provided we adorn the brane with additional charges.
In particular, we can consider the bound state with a longitudinal type IIA
string. In the matrix string language, the winding number $m$ of this string
is identified with the total quantized world-sheet momentum $P = \oint T_{01}$
of the matrix string
\be
\label{P}
P = 
\oint \! d\sigma \, \tr\Bigl(\Pi_iDX^i + \theta^T D
\theta\Bigr) = m. 
\ee
In addition, we will assume that this bound state of
NS fivebranes and type IIA strings 
carries a non-zero longitudinal momentum $p_9$ per unit length.

In the standard large limit of matrix theory \cite{banks}, the 
momentum $p_+$ is represented by the ratio $p_+ = N/R_9$ in a limit
where both $N$ and $R_9$ tend to infinity, keeping their ratio fixed.
Here however we have to take a different limit that leads to a finite
longitudinal momentum density 
\be
{p_+\over R_9}={N\over R_9^2}.
\ee 
Clearly, this implies that in the large $N$ limit the radius $R_9$
necessarily scales as $\sqrt N$. A similar limit was considered in
\cite{aharony-berkooz} in the study of longitudinal membranes in
matrix theory\footnote{In our context these configurations are indeed
the longitudinal type IIA strings.}.

In the presence of the black string configuration, the unbroken
part of the supersymmetry algebra takes the form
\be
\{Q^\a,Q^\b\} = (p_-\id + w^+ \gamma^9  +  W^+_{ijkl} \gamma^{ijkl})^{\a\b}
\label{lcalgebra}
\ee
with the finite central charge densities (in string units $\alpha'=1$)
\be
\label{charged}
{w^+\over R_9} = m, \qquad \quad {W^+_{ijkl} \over R_9} = k {V\over g_s^2} 
\epsilon_{ijkl}.
\ee
where $V$ is the volume of the four-torus.
The extremal black string is annihilated by the
right-hand side of (\ref{lcalgebra}). From this one derives the
following expression for the black string tension 
\be
{M_{ext}\over R_9} = 
{N\over R_9^2} + {m} + {kV\over g_s^2}.
\ee
The last term represents the contribution of the fivebrane with
tension $1/g_s^2$ wrapped $k$ times around the four-torus.
This characteristic dependence on $g_s$ of the mass of the 
solitonic NS fivebrane can be read off immediately from the 
Hamiltonian (\ref{H}), using the definitions (\ref{wplus})
and (\ref{charged}).

\newsubsection{The extremal 5D black hole}

Upon compactification of the spatial light-cone direction $x^9$ 
on a circle of radius $R_9$, the six-dimensional black string 
described above becomes a five-dimensional black hole with 
quantized momentum 
$$
p_9 = {N\over R_9}.
$$ 
In the extremal case, this black hole has a mass given by
\be
\label{extmass}
M_{ext} = 
{N\over R_9} + {mR_9} + {kR_9V\over g_s^2},
\ee
and is directly related
through a sequence of dualities to the five-dimensional type IIB
black hole solution considered in \cite{vafa-strominger}. 

Let us briefly pause to recall the intermediate steps of this duality mapping.
Starting on the type IIA side, we apply a T-duality along the
compactified $x^9$ direction mapping the type IIA string on a type IIB
string, and interchanging winding number and momentum.  Under this
T-duality the IIA NS-fivebrane becomes the IIB NS-fivebrane.  Next,
via the strong-weak coupling symmetry of the type IIB theory, we can
turn this into a configuration of $N$ D-strings in the presence of $k$
D-fivebranes.  Finally, by a complete T-duality along all 4 directions
of $T^4$ we obtain $N$ D-fivebranes and a D-string with winding number
$k$. Hence, through this duality sequence, the NS-fivebrane eventually
becomes a D-string, while the original type IIA string has become the
D-fivebrane. As a result, the matrix-valued fields $X^i$ on the matrix
string are identified with the covariant derivatives $D_i$ in the
$U(N)$ SYM theory on the world-volume of the D-fivebrane, while the
D-strings represent the YM instanton configurations.  Tracing this back 
to the type IIA language, one obtains the soliton description of the 
NS-fivebrane \cite{ganor,branes} given above.

For large values of the given 
charges $(k, m, N)$, this extremal black hole is well-known to have a 
macroscopically large horizon area and corresponding Bekenstein-Hawking 
entropy, given by 
\be
\label{entropy}
S_{BH}= 2\pi \sqrt{Nmk}.
\ee
A microscopic interpretation of this formula for the case of the type IIB RR 
black hole  is obtained \cite{vafa-strominger} by considering all possible 
quantum fluctuations of a multiply wound D-string bound to a D-fivebrane. 

For BPS-configurations, it turns out that this compactification of 
the 9-th dimension (which turns the light-cone plane into a cylinder 
$S^1\times \R$) can be achieved by considering the matrix string model 
with finite $N$, which becomes identified with the discrete 9-momentum. 
It should be emphasized, however, that this procedure
inevitably falls short of describing the non-BPS dynamics of the 
theory, as will become clear below.

The winding number $m$ around the compactified circle $R_9$ is incorporated 
in the matrix 
string theory via an appropriate modification of the mass shell and level 
matching conditions, which are dictated by the space-time
supersymmetry algebra (see appendix B).  The level matching condition 
equates the total world sheet momentum of the matrix string to 
the integer winding number $m$ around the $R_9$ direction,
while the mass-shell relation takes the form
\be 
N H =  p_0^2 - p_9^2 - w_9^2.
\ee
For BPS states, this exact structure indeed arises from the matrix 
string theory at finite $N$ \cite{matrix-string}. Let us explain how
this goes.

In the IR limit of the 1+1-dimensional SYM theory, the model enters into
the Higgs phase. The potential forces the matrix coordinate fields $X^i$ 
to mutually commute, so that they can be simultaneously diagonalized.
The invariance under the remaining discrete gauge rotations 
tells us that the eigenvalues take value on the symmetric 
space $S^N\R^8$. The various twisted sectors of this orbifold CFT are labeled
by partitions $\{n_i\}$ of $N$, 
\be
\sum_i n_i = N,
\ee
where the individual strings have momentum $p^{(i)}_9=n_i/R_9$. The total 
worldsheet energy-momentum in this twisted sector decomposes as 
$L_0^{tot}= \sum_i {1\over n_i} L_0^{(i)}$.
The $S_N$ orbifold construction naturally
implements the individual level matching relations \cite{matrix-string}  
\be 
L_0^{(i)} -
\Lbar_0^{(i)} = n_i m_i = p_9^{(i)} w_9^{(i)}.
\ee 
It is therefore indeed consistent to interpret $n_i$ and 
$m_i$  as the quantized momentum and winding of the individual 
strings in the $9$-direction. Note, however, that this description
only works within the BPS sector, since the individual discrete string
momenta $n_i$ are all required to be positive. 

We would now like to use this construction to consider the  BPS bound states 
of $k$ NS fivebranes and  $m$ IIA strings in the compactified 
light-cone situation. In this situation the central charges
that appear in the unbroken supersymmetry algebra
\be
\{Q^\a,Q^\b\} = (p_-\id + w_9 \gamma^9  +  W_{9ijkl} \gamma^{ijkl})^{\a\b}
\label{algebra}
\ee
are finite quantities. The BPS condition
\be
\label{condition}
\epsilon^\a Q_\a  |{\mbox{\sc bps}}
\rangle = 0
\ee
is compatible with this algebra, provided the spinor $\epsilon$
satisfies
\be
( p_- + w_9 \gamma^9  +  W_{ijkl9} \gamma^{ijkl} )\epsilon = 0.
\ee
This makes $\epsilon$ eigenspinor of both $\gamma^9$ and
$\gamma^{ijkl}$. 
%\be (1-\gamma^9)(1-\Gamma)(\gamma^9 E 
%+\gamma^{9i} DX_i + \gamma^i\Pi_i + \gamma^{ij}[X_i,X_j]) = 0
%\ee
For a fivebrane wrapped $k$ times around the torus $T^4\times S^1$
we have $W_{ijkl9}= k \, R_9 V \epsilon_{ijkl}$,
where the epsilon symbol is short hand for the unit
volume element of the $T^4$.

The resulting BPS equation of motion is obtained by requiring
that the supersymmetry variation of the fermions vanishes
\be
\delta_\epsilon\theta = \left(\gamma^9E + \gamma^i\Pi_i + 
\gamma^{i9}DX_i + \gamma^{ij}[X_i,X_j]\right)\epsilon=0
\ee
for the parameters $\epsilon$ determined above. This gives the equations
\be
\label{sd}
[X_i,X_j] \, = \, \epsilon_{ijkl} \, [X^k,X^l],
\ee
and
\ba
\label{soliton}
DX_i \, = & \!\! \Pi_i , \qquad  \ \qquad  
[X_i,X_I] \, \! & =\, 0, \\[2mm]
 E  \, = & \!\!  0,  \qquad \qquad \; [X_I,X_J]  \! & =\, 0.
\label{flat} 
\ea 
Here $i,j,k,l$ denote only the internal $T^4$ directions, while
$I, J$ denote the transversal uncompactified dimensions.  

Via the identification of $X^i$ with a covariant derivative on the dual $T^4$,
we recognize in equation (\ref{sd}) the self-duality equation of the

corresponding gauge field configuration. The solutions of the
fivebrane equation (\ref{soliton}) are thus in one-to-one
correspondence with maps from $S^1 \times \R$ (with $\R$ the time
direction) into the space of self-dual Yang-Mills instantons on
$T^4$. The  equation (\ref{soliton}) requires that these maps are
holomorphic.  Finally, the last two equations (\ref{flat}) require
that the four scalar fields $X^I$, that describe the four transverse
coordinates to the fivebrane, mutually commute and be covariantly
constant on the $T^4$. For generic instantons, this condition leaves
only the constant matrices $X^I = x^I_{cm} \id$ that define the center
of mass of the fivebrane.

\newsubsection{Recovering the world-volume theory}

We thus find that the space of collective coordinates of the wrapped
NS-fivebrane in type IIA theory is identical to the moduli space of
self-dual Yang-Mills instantons on $T^4$. The $k$-instanton moduli
space of the $U(N)$ theory is a 4$Nk$ dimensional space. It is
conjectured to be described by a hyperk\"ahler deformation of the
symmetric product space $S^{Nk}\,T^4$ \cite{ama}. Note that this
symmetric space only depends on the product $Nk$ and in particular is
symmetric under the exchange of $k$ and $N$.

One way of understanding this fact mathematically is through the
Nahm-Mukai transformation (see {\it e.g.} \cite{mukai,Pierre}) that
maps the $k$ instanton moduli space of $U(N)$ Yang-Mills theory on
$T^4$ to the $N$ instanton moduli space of $U(k)$ theory. This Nahm
transformation can be regarded as a low-energy manifestation of the
maximal $T$-duality on $T^4$, since it interchanges the 5-brane number
(= rank $N$) with the 1-brane number (= instanton number $k$).
Indeed, T-duality considerations for D-brane bound states have led
precisely to the identification of the instanton moduli space with a
symmetric product \cite{vafa}.  For the special case $k=1$, \ie for a
singly wrapped fivebrane, the Nahm transformation relates the one
instanton moduli space to the configuration space of $N$ point-like
abelian instantons, or actually torsion free sheaves \cite{mukai},
whose positions are identified with the coordinates on $S^{N} T^4 $.
For this case the space $S^{N}T^4$ is therefore just the configuration
space describing the transversal positions of the strings on the
fivebrane.
 
The fact that this moduli space has the structure of the symmetric 
product $S^NT^4$ is known to have important consequences
\cite{vafa-strominger,maldacena-susskind,orbifold}. Namely, it tells
us, via the same arguments as used in the discussion of the free
matrix string theory in \cite{matrix-string}, that the sigma model can 
be interpreted as a second-quantized string theory living in $5+1$
dimensions.  From the M-theory perspective these are strings that are
confined to the five-brane world-volume. In this way we make contact with
the original analysis of \cite{vafa-strominger} and with our 
papers \cite{letter,5brane}, where one can find a
detailed discussion of these strings and explicit formulae for the
BPS masses and spectra, etc. 

As indicated in \figuurplus{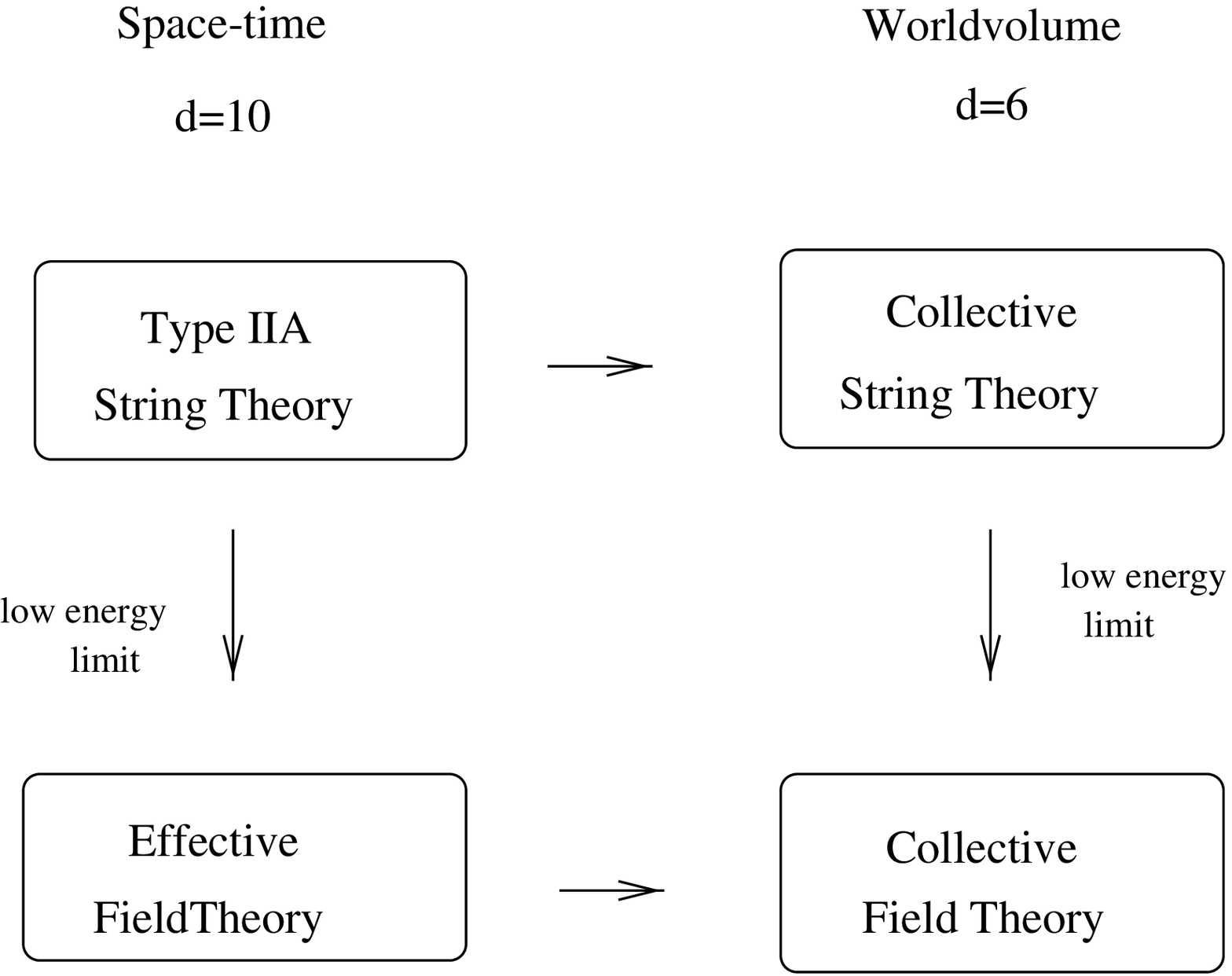}{9cm} {The effective world-volume
dynamics of the solitonic NS fivebrane of type IIA string theory is
described by a $d=6$ ``collective'' string theory, whose low-energy
degrees of freedom coincide with the usual collective fields.} the
effective world-volume description of the solitonic NS fivebrane is
usually derived by considering the collective modes that one deduces
from the low energy effective field theory \cite{fivebrane}. In this
paper we have in some sense repeated this procedure directly at the
string level, and thus identified a collective string theory that
describes the fivebrane fluctuations.  Here we briefly recall how the
string degrees of freedom indeed reduce to the usual collective modes
of the fivebrane in the low energy limit, for more details
see \cite{letter,5brane}.

The world-volume string of the fivebrane has half the world-sheet
degrees of freedom of the type IIB superstring. In a light-cone gauge,
there are 4 transverse bosonic fields $\xX$ inside the world-volume of
the fivebrane.  The left-moving string world-sheet fields
are\footnote{ Here we use chiral spinor indices $a$ and $\dot a$ of
the transversal $SO(4)$ rotation group, and the indices $\alpha$
($\dot\alpha$) are (anti-)chiral spinor indices of another $SO(4)$
identified with space-time rotations, which on the fivebrane
worldvolume are realized as an R-symmetry.} $(\xX^{a\dot{a}},\
\sS^a_\alpha)$ and the right-movers are $(\xX^{a\dot{a}},\
\sSbar_{\dot\alpha}^a)$.  The ground states must form a multiplet of
the left-moving zero-mode algebra
$\{\sS_a^\alpha,\sS_b^\beta\}=\epsilon_{ab}\epsilon^{\alpha\beta}$,
which gives 2 left-moving bosonic ground states $|\alpha\rangle$ and 2
fermionic states $|a\rangle$. By taking the tensor product with the
right-moving vacua one obtains in total $16$ ground states
representing the massless tensor multiplet on the five-brane.
Specifically, we have states $|\alpha\dot\beta\rangle$ that describe
four scalars $X^{\alpha\dot\beta}=\sigma_i^{\alpha\dot\beta} X^i$
transforming a vector of the $SO(4)$ R-symmetry, while $|\alpha
b\rangle$ and $\psi^\alpha$ and $\psi^{\dot\alpha}$.  Finally, the
RR-like states $|ab\rangle$ decompose into a fifth scalar, which we
call $Y$, and the 3 helicity states of the self-dual tensor field.
These are indeed the fields that parametrize the collective
excitations of the five-brane soliton.

Let us make some speculative remarks about the possible interactions
of these 5+1 dimensional strings. For simplicity we take $k=1$ and
consider one fivebrane. In the matrix formulation of the fundamental
type IIA string as a large $N$ limit of the $S^N\R^8$ orbifold CFT, we
have shown in \cite{matrix-string} that the perturbative joining and
splitting interactions are reproduced by deforming the CFT by means of
an appropriate $\Z_2$ twist field $V_{int}$. In that case, this
interaction vertex has total dimension 3 and therefore represents an
irrelevant operator in the CFT that disappears in the IR weak
string-coupling limit. In the case of the $c=6$ string theory on the
fivebrane described by the $S^NT^4$ orbifold, the analogous $\Z_2$
twist operator has total dimension 2 and hence is a marginal operator.
So, at least to first order in perturbation theory, turning on the
string interactions will preserve the conformal invariance of the
second quantized sigma model. In fact, this particular $\Z_2$ twist
field is well-known to be the blow-up modulus of the ALE space
$\R^4/\Z_2$, describing the local $A_1$ type singularity.  Similarly,
the higher order twist $\Z_n$ fields that are produced by contact
terms, will serve to blow up the $A_{n-1}$ singularities of the
symmetric product. An obvious guess is that the blown-up CFT that
represents the interacting world-volume string theory is a sigma model
on the large $N$ Hilbert scheme of $T^4$, which is a distinguished
hyperk\"ahler resolution of the symmetric product.

\newcommand{\qq}{{\bf q}}

\newsubsection{Towards duality invariance}

An attractive feature of matrix theory is that it makes the
geometrical M-theory symmetries that act on the transversal space
manifest, and that most of these transformations are non-perturbative
dualities in the IIA interpretation. The perturbative IIA T-duality
symmetries, on the other hand, have unfortunately become more
obscure. Various T-duality symmetries have been identified in matrix
theory \cite{ganor,susskind,rozali} and generally involve
non-perturbative dualities of the SYM model.  In the toroidal
compactification to five dimensions, the full string duality group is
$E_{6(6)}(\Z)$. Of this large group, only an $SL(5,\Z)$ subgroup is a
manifest symmetry of the 5+1-dimensional SYM theory on $T^5$.
Furthermore, of these transformations only the $SL(4,\Z)$ subgroup
that acts on the transversal four torus $T^4$ is part of the IIA
T-duality group $SO(5,5;\Z)$. Below we indicate how some of these
extra duality symmetries can be incorporated in the present framework.

To begin with, let us include all possible charges by allowing
for non-zero fluxes and topological charges in the 5+1-dimensional 
SYM model. These fluxes are most conveniently identified as 
the various possible RR-charges and NS-charges of the type IIB 
string theory. In this case the D-branes have odd dimensions, and
a general D-brane configuration wound around the five-torus 
is therefore characterized by a 16 component vector in the odd 
integral homology 
\be
\label{config}
\qq \in H_*^{odd}(T^5,\Z).
\ee
This vector $\qq$ transforms in a spinor representation of 
the $SO(5,5;\Z)$ T-duality group.
In the SYM terminology, these charges arise as the rank $N$,
and the magnetic fluxes and instanton winding numbers defined by
\ba
m_{ij} \is \int \tr F_{ij}, \nonu
k^m \epsilon_{mijkl} \is \int \tr F_{[ij} F_{kl]},
\ea
which respectively count the D fivebrane, D 3-brane, and
D-string winding numbers of the type IIB theory.

The total number of charges in 5-dimensional type II string theory is 27.
Ten of the remaining 11 NS charges correspond to the KK momenta 
and IIB string winding numbers, which arise in the SYM language 
via the integrated momentum flux $p_i = \int T_{0i}$ and 
the electric flux $e_i$,
\ba
\label{moment}
p_i \is \int \tr( E^j F_{ji} + \Pi^I D_i X_I + \theta^T D_i \theta), \\[3mm]
e_i \is \int \tr E_i.
\ea
The one charge that at this point seems to be missing
in the matrix formalism is the NS fivebrane charge $q_5$. We are therefore
forced to put it to zero for the moment. 

Via the above translation code, the effective stringlike instantons 
identified in the previous subsection, exactly correspond to 
the D-brane ``intersection strings'' of \cite{vafa-strominger}.  
For a general D-brane configuration (\ref{config}), the
total winding number of the intersection strings in given by the
intersection pairing
\be
\half (\qq\wedge \qq)_i = N k_i + \half (m \wedge m)_i.
\ee
This is indeed also the total winding number of the stringlike
instantons in the SYM theory. Crucial for this interpretation is the
fact that the moduli space of instanton configurations on $T^4$ with
rank $N$, instanton number $ch_2=k$, and fluxes $c_1=m$ is again related to
a symmetric product space $S^nT^4$, now with $n=Nk +\half m\wedge m$. 
In the case of instantons on the five-torus, the integer $n$ will be
the integer length of the vector $(\qq \wedge \qq)_i$, relative to the 
primitive lattice vector in the same direction. The length is equal to
the g.c.d.\ of the five components of $\half \qq\wedge \qq$. 

The BPS multiplicities are unique functions of the macroscopic 
quantum numbers defined above. Their asymptotic growth for
large charges is predicted by the Bekenstein-Hawking entropy
formula. Translating the $E_{6(6)}$ invariant entropy formula
\cite{letter} into the above notation, this macroscopic entropy
formula takes the form
\be
\label{sbh}
S_{BH} = {2\pi}\sqrt{\Bigl(Nk_i + \half(m\wedge m)_i\Bigr)
\Bigl(p_i -e^j m_{ij}/N\Bigr)}
\ee
This formula is obtained by putting one of the 27 charges to zero,
which breaks the duality group down to $SO(5,5;\Z)$, which is the
IIB T-duality group.
As discussed in our previous papers, this result is reproduced 
via the effective string description
by matching the total winding number and momentum of the individual
strings with the above expression in terms of the macroscopic 
quantum numbers.  Introduce occupation numbers 
$N^I_{n,\tilde{n}}$ that keep track of the number of individual strings 
$|I,n,\tilde{n}\rangle$ with winding number $\tilde{n}$ and momentum $n$.
Then we have 
\be
\label{een}
\sum \tilde{n}_i N^I_{n,\tilde{n}} = N k_i + \half (m \wedge m)_i
\ee
while the definition (\ref{moment}) of 
$p_i$ as the total integrated momentum
gives that
\be
\label{twee}
\sum n_i N^I_{n,\tilde{n}} = p_i -  {e^j m_{ij}\over N}.
\ee
Here the second term on the right-hand side represents the
zero-mode contribution to (\ref{moment}). The BPS condition
further implies that all momenta and windings of the individual 
strings must be aligned in the same direction. Given that the
effective BPS strings are non-interacting, and can oscillate in 
4 transverse directions, it is now a standard computation to 
reproduce from (\ref{een})-(\ref{twee}) the $SO(5,5;\Z)$ formula 
(\ref{sbh}) for the entropy.

\newsubsection{Non-extremal black strings}

Up to now we have concentrated on the extremal BPS solution. To which
extent can we relax the BPS condition and study the more generic
non-extremal cases? As we have already noticed, compactification of
the longitudinal fivebrane via a finite $N$ matrix model is only valid
within the BPS sector. So to go off extremality we are forced to again
decompactify the light-cone, and consider the six-dimensional black
string with a finite longitudinal momentum density discussed before.
Apart from the fact that this object is infinitely stretched in one
direction, its geometry and many of its other physical properties are
those of the five-dimensional black hole solution, which are
summarized in Appendix A.  Extensive quantities as the energy and
entropy are now measured per unit length.  In matrix theory these
stretched black holes appear in the limit where the compactification
radius is taken to infinity as $R_9 \sim \sqrt N$.

Detailed comparisons with the expected thermodynamic properties of
near-extremal 5-dimensional black holes have indicated that the
description in terms of collective string degrees of freedom that live
on the fivebrane world-volume extends beyond the BPS regime
\cite{near,blackhole2}. We would now like to show how this description
arises in matrix theory\footnote{We thank F. Larsen and J. Maldacena
for valuable discussions about the D-brane description of non-extremal
black holes.}.

We return again to the situation in which only three charges $N$, $m$
and $k$ are non-vanishing.  Recall the description in section 4 in
terms of the effective world-volume string in terms of the sigma model
on the $k$ instanton moduli space $S^{Nk}T^4$.  One way to go
off-extremality is to relax this chirality condition, so that in the
instanton string sigma model on $S^{Nk} T^4$ we will now allow for
both left-movers and right-movers. The effect of this has been studied
extensively in the D-brane black hole literature, and leads to
separate quantities $L_0=m_L$ and $\Lbar_0= m_R$, that indicate the
total oscillation numbers of the left-moving and right-moving modes in
the world-sheet conformal field theory. The level matching relation
$L_0-\Lbar_0=m$ that relates the difference $m_L-m_R$ to the total
winding number is still satisfied, but the individual winding numbers
$m_i$ of the strings are no longer all of the same sign. Notice
further that the left and right oscilation levels $m_L$ and $m_R$ are
in general not integers, but are quantized only in units of $1/kN$.

The fact that we allow for string excitations of both chiralities 
translates into an extra parameter $\alpha$ that appears in the non-extremal 
black string geometry, via  $m_R = e^{-4\alpha} m_L$ (see Appendix A). The 
Bekenstein-Hawking formula predicts that in this particular non-extremal 
situation, the relevant black string degrees of freedom are counted by 
the entropy density 
\be
S_{BH} = 2\pi \sqrt{N k} \left(\sqrt{m_L} + \sqrt{m_R}\right) 
\ee
This result appears naturally from the effective string ensemble by the standard 
formula for the asymptotic growth of the number of states as a function of
the oscillator level. In the large $N$ limit, the entropy density $S/R$ 
approaches a finite value.

A useful remark at this point is that, as discussed in the previous
section, the description of the non-chiral effective string has a 
T-duality symmetry that relates the quantum number $m$ with the 
combination $Nk$. While this is not immediately evident from orbifold
sigma model discription, it follows quite directly from the discussion
in the preceding section leading to equations (\ref{een}) and (\ref{twee}).
Indeed, there is another way to obtain a non-extremal black string configuration,
which can be viewed as the T-dual of the situation just described.
If we drop the BPS restriction, it is clear the restriction to self-dual
Yang-Mills configurations must be relaxed. We must therefore allow for 
instantons as well as anti-instantons of the 5+1 dimensional SYM model. 
By duality, the statistical description in this sector can be obtained
from the one given above by interchanging the role of the two level matching 
relations (\ref{een}) and (\ref{twee}). We will return to this argument
in a moment.

In the near extremal situation, where we can make reliable 
comparisons with the D-brane computations, the SYM theory is weakly coupled. 
We expect therefore that  in this regime the dynamics can be reliably described  
in a dilute instanton gas approximation, in terms of a collection of 
$k_L$ instanton strings and $k_R$ anti-instanton strings. In the type IIA 
theory these correspond respectively to fivebranes and anti-fivebranes, so  
that the total instanton charge $k_L-k_R=k$ is identified with the total 
fivebrane wrapping number. The weak coupling limit of the IIA theory 
arises by stretching one directions $S^1$ (that in M-theory is given by
$1/R_{11}$) of the five-torus. Physically this implies that the
instanton strings become oriented in this direction. 

An adiabatic argument \'a la \cite{vafa-strominger} suggests that the 
relevant SYM field configurations behave like small instantons in the 
transversal $T^4$, whose positions are slowly varying as a function 
of the world-sheet coordinate $\sigma$. In this way we again recover 
a sigma-model with target space the moduli space of semi-classical SYM 
configurations.
If we allow ourselves to use the dilute instanton gas picture, this moduli
space again looks like a symmetric product, but   now of a factorized form
\be
\label{sym}
S^{Nk_L}T^4 \times S^{Nk_R}T^4.
\ee
The corresponding effective strings thus appear with both positive and 
negative orientation. 

As with the left- right oscillation levels $m_L$ and $m_R$, the appearance 
of the anti-instantons translates into an extra parameter $\gamma$ of the 
non-extremal black string geometry, via  $k_R = e^{-4\gamma} k_L$ (see 
Appendix A). We can go off-extremality in this direction, while keeping 
$m_L=m$ and $m_R=0$, so that the strings have only chiral excitations.
In this case the Bekenstein-Hawking formula predicts the following 
result for the black string entropy 
\be
S_{BH} = 2\pi \sqrt{N m} \left(\sqrt{k_L} + \sqrt{k_R}\right) 
\label{nonext2}
\ee
Our next challenge is to derive this expression from this present
formalism.  At a first glance, this result for the entropy appears to
follow quite directly as the asymptotic number of states of the sigma
model on the space (\ref{sym}), by the standard relation the central
charge to the asymptotic growth of the number of states. However, one
encounters a subtlety at this point, since to arrive at
(\ref{nonext2}) one must consider states for which the two types of
strings each have the same total oscillation number equal to $m$. A
priori, however, one would be inclined to allow for different
oscillation numbers, and interprete $m$ as the sum of the two. Let us
explain how this prescription indeed follows as a necessary
consequence of the T-duality of the effective string theory.

To this end, let us return to the non-extremal sector with only
instanton strings ($k_R=0$) but with non-chiral excitations
($m_R\not=0$), and see how it could be described in a T-dual
fashion. Indeed the description given above in terms of the left- and
right-moving excitation numbers $m_L$ and $m_R$ in the orbifold CFT on
$S^{Nk}T^4$ breaks the T-duality symmetry between $m$ and $Nk$.
Interchanging the role of the winding and momentum quantum numbers, we
can equally well encode the multi-string degrees of freedom in terms
of a dual orbifold sigma model, where $m$ now determines the order of
the symmetric product and $Nk$ gives the oscillation level.  

Going
off-extremality by introducing $m_L$ and $m_R$ translates into a
modification of the dual sigma model. Namely, in this dual picture the
contributions from the individual strings to $m$ have a geometric
interpretation in terms of the length of the permutation cycles
labeling the twisted sectors. If we want to be able to separate this
into a left-moving and right-moving contributions adding up to $m_L$
and $m_R$, we are led to introduce separate orbifold target spaces
$S^{m_L} T^4 \times S^{m_R} T^4$, which each describe multiple
strings.  The states of the original $S^{Nk} T^4$ orbifold can now be
identified with the chiral states in the two factors $S^{m_L} T^4$ and
$S^{m_R}T^4$, each with the {\it same} oscillation level $L_0^{dual} =
Nk$. While this dual description may seem an artificial way of
incorporating two chiralities of the instanton strings, after applying
the T-duality we reproduce precisely the picture sketched above of a
gas of chiral instanton and anti-instanton strings described by the
product sigma model (\ref{sym}).

Although this T-duality gives an argument for the curious ``level
matching'' condition that equates the oscillation level of the two
types of strings, the precise geometrical justification is not clear
to us. Presumably a more careful understanding of the interaction
between the instanton strings and anti-instanton strings can shed some
light on this. If however one accepts this description, it seems that via
a straightforward generalisation one can include both kinds of
non-extremality with $k_R$ and $m_R$ non-zero and reproduce the entropy
formula (\ref{ent}) by including both left- and right-moves on the 
product sigma model.

This description of the microscopic fivebrane degrees of freedom in
principle allows one to study the near-extremal black string dynamics,
such as the excitation spectrum, absorption and emission processes
\cite{dasmathur,juandy}, traveling waves \cite{gary}, etc.  It is
important to note that in the decompactification limit $R_9
\rightarrow \infty$, the near-extremal black string energy spectrum
becomes continuous, since it allows for arbitrarily long wavelength
excitations.
It is worth pointing out that in this formulation
the tension of such a non-extremal black string is reproduced as well.
We have
\be
{M\over R_9} = {p_+\over R_9} + {p_-\over R_9} = {N\over R_9^2} + H
\ee
with $H$ the matrix string Hamiltonian (\ref{H}). In the above 
effective description the Hamiltonian receives two kind of contributions:
the (anti)instanton configurations give rise to a potential energy
term equal to 
\be
{1\over g_s^2} \oint \tr [X^i,X^j]^2 = {V\over g_s^2} (k_L + k_R).
\ee
The excitations of the effective strings produce the energy
\be
\oint \tr \Bigl(\Pi_i^2 + (DX_i)^2\Bigr) = L_0 + \Lbar_0 = m_L + m_R
\ee
Taken together this gives the non-extremal mass formula (\ref{mass}).
Note that the T-duality of the effective strings that we used above, is
not a symmetry of the space-time properties of the black string.

\newsubsection{Recovering the space-time geometry}

We have seen that the matrix setup is quite well adapted to
extract all the effective degrees of freedom that describe the type
IIA fivebrane and its space-time manifestation as a black string in
5+1-dimensions. The description that we obtained is in direct
accordance with known results, but one might hope that the present
context of matrix theory opens up new ways of examining the dynamical
interactions of these black strings with their environment. In the
following subsection we will indicate how the fivebrane in particular
influences the propagation of the fundamental type IIA strings in its
neighbourhood.

The matrix theory description of the fivebrane is rather different
from its description as a soliton of the low energy effective field
theory. Nevertheless in the weak coupling limit $g_s\ra 0$ one would
like to be able to recover the classical fivebrane geometry including
its anti-symmetric tensor field from the matrix theory. To study this
question we will introduce a probe in the matrix theory, namely an
additional type II string that moves in the background of the
fivebrane. We will argue that the effective
world-sheet action of the string probe, derived from matrix theory,
indeed coincides with the known sigma model for the fivebrane soliton
\cite{callanetal}.

In matrix theory both the string probe as well as the fivebrane
background will be described in terms of the same $N\times N$ matrix
coordinates $X_i$.  The purpose of the probe is to test the geometry
of the fivebrane and therefore we would like to introduce it in such a
way that it has a minimal effect on the other degrees of freedom of
the matrix theory. To achieve this we will add one row to the matrix
coordinates $X^i$, so that we are dealing with a $U(N+1)$ matrix
theory.  Next we explicitly split off the $U(N)$ part $Z_i$, that
describes the YM instanton configurations carrying the fivebrane
charge, by writing 
\be 
X_i \rightarrow \twomatrix{Z_i}{Y_i}{\overline{Y}_i}{x_i} 
\label{ZYx}
\ee 
Here the matrix entry $x_i$
represents the coordinate of a propagating type IIA
string (with infinitesimal light-cone momentum $p_+$). 
Similarly, we can write 
\be 
\theta_\a \rightarrow
\twomatrix{\zeta_\a}{\eta_\a}{\overline{\eta}_\a}{\theta_\a} 
\ee 
This
matrix theory description of the string probe and fivebrane is closely
related to the recently studied case of a D-string probe in the
background of a D-fivebrane \cite{dstring-probe,wadia}.  In the D-brane
formalism the dynamics of the D-string probe is obtained by taking
into account its interaction with the 1-5 open strings that connect
the probe with the fivebrane. In \cite{dstring-probe} it is 
shown that by integrating out these extra degrees of freedom one
obtains an explicit form of the effective metric that describes the
propagation of the D-string probe. In the matrix setup,
the analogue of the 1-5 string 
degrees of freedom that mediate the interaction between the black hole
and the probe string  are the off-diagonal $1\times N$ 
components $(Y,\overline{Y})$.

The idea of the computation is now to fix $Z_i$ to be some given
instanton configuration on $T^4$ plus a diagonal part $z_I {\bf
1}_{N\times N}$ that describes the location of fivebrane in the
uncompactified space. We can then derive an effective action for the
$x$-part of the $U(N)$ matrix by integrating out the off-diagonal
components $Y$ and $\eta$.  For simplicity, we will first consider
only the center of mass coordinate $x_I$ of the probe string. 
We will later study the finite $\alpha'$ corrections.
The leading contribution comes from the constant
zero-modes in the fivebrane background, that satisfy 
\be
Z_{[i} Y_{j]} = 0, \qquad \quad \gamma^i_{\a\b} Z_i \eta^\beta = 0.
\ee
Since $Z_i$ corresponds to a $k$-instanton configuration we can use 
the index theorem to determine that there are $k$ such zero-modes,
both for the $Y$ and $\eta$ field.
The relevant quadratic part of the action takes the form
\ba
S \is \int \! d^2\!\sigma \; \Bigl(\, (D {Y}_i)^2 +
|x_I\!-\! z_{I}|^2 | Y_{i}|^2\, \Bigr) \nonu
&& \qquad  + \; \int d^2\!\sigma\; 
\Bigl( \, \overline{\eta}_\a^T D_+\eta_\a +
\overline{\eta}_{\dot\a}^T D_-\eta_{\dot\a} + \overline{\eta}_\a^T
\gamma_{\a\dot\b}^I (x\! -\! z)_I )\, \eta_{\dot\b} \; \Bigr) 
\label{off}
\ea
Here we suppressed the summation over the index labeling the 
$k$ zero-modes.

Let us explain the fermion chiralities as they appear in this
Lagrangian.  In the original matrix string action the $SO(8)$
space-time chiralities are directly correlated with
the world-sheet chirality, reflecting the type IIA string setup,
giving left-moving fermions $\theta^\a$ and right-moving fermions
$\theta^{\dot\a}$.  The $T^4$ compactification breaks the $SO(8)$
into an internal $SO(4)$ that acts inside the four-torus times 
the transversal $SO(4)$ rotations in space-time. The
spinors decompose correspondingly as ${\bf 8}_s \ra ({\bf 2}_+,{\bf
2}_+) \oplus ({\bf 2}_-,{\bf 2}_-)$ and ${\bf
8}_c \ra ({\bf 2}_+,{\bf 2}_-) \oplus ({\bf 2}_-,{\bf 2}_+)$. 
The instanton on $T^4$ with positive charge $k>0$ allows only
zero modes of positive chirality, selecting the spinors that transform
as ${\bf 2}_+$ of the first $SO(4)$ factor. So, through this process the
space-time chirality of the fermion zero modes gets correlated with
the world-sheet handedness, giving $k$ left-moving fields $\eta_L^\a$
and $k$ right-moving fields $\eta_R^{\dot\a}$, where $\a,\dot\a=1,2$
now label the space-time spinors.

Following the calculation of \cite{dstring-probe} we can now
evaluate the determinants, which yields the one-loop contribution 
to the string world-sheet lagrangian 
\be
k \, {(D x)^2 \over |x - z|^2}.
\ee
In combination with the classical contribution, this leads to a sigma-model
action for a string moving in a transversal background metric
\be
ds^2 = \Bigl(\, 1+ {k\over |x-z|^2} \,\Bigr) (dx)^2 
\ee
which is the well-known geometry of the NS fivebrane as seen in leading
order in $\alpha'$ by  the type IIA strings \cite{callanetal}.

Most characteristic of the solitonic fivebrane solution is the fact
that it carries magnetic charge with respect to the three-form field
strength $H = dB$ of the NS anti-symmetric tensor field $B$, that is
\be 
\label{H3}
\int_{S_3} H = k 
\ee 
for a three sphere enclosing the brane. This
result can be recovered from the matrix language by a direct analogous
calculation as above. Essentially all we need is a simple adaptation
of the analysis of \cite{berkooz-douglas}, where the effect of the
fivebrane background was incorporated in matrix theory by adding an
appropriate hypermultiplet to the membrane action. As emphasized
above, these additional hypermultiplets indeed arise automatically as
the off-diagonal fields $Y,\eta$.  

To recover the background anti-symmetric tensor field $B$, one must
necessarily extend the probe degrees of freedom to represent an
extended type IIA string world-sheet $x^I(\sigma,\tau)$ and not just
its center of mass $x^I(\tau)$, as we did in the above computation. We
will now show how integrating out the $Y,\eta$ fields in this more
general context leads to the well-known CFT of \cite{callanetal} that
describes the solitonic fivebrane geometry to all orders in $\a'$.
Our computation is valid for $x^I \approx z^I$, which corresponds to
the throat region very close to the black string horizon\footnote{We
thank L. Alvarez-Gaum\'e for suggesting the following argument.}.

First we note that the transversal rotation group $SO(4)\simeq
SU(2)_L\times SU(2)_R$ acts on the relative coordinate field
$(x-z)^I$ via left and right multiplication
\be
\gamma_I (x-z)^I  \ra g_L \cdot \gamma_I (x-z)^I \cdot g_R^{-1}.
\ee
This group action can be used to 
decompose the $2\times 2$ matrix $\gamma^{\a\dot\b}_I(x-z)^I$ 
in terms of a radial scalar field $\varphi$ and
group variables $g_L,g_R$ as 
\be
\gamma_I (x-z)^I = e^\varphi g_L g_R^{-1}.
\ee
The horizon is located at $r= e^\varphi = 0$. 
The combination $g=g_Lg_R^{-1}$ labels the
$SO(3)$ angular coordinate in the transversal 4-plane.
Notice that the above relation allows gauge transformations 
$(g_L,g_R) \ra (g_L h,g_R h)$ with $h\in SU(2)$.

Inserting the above expression for $x^I$ into the action (\ref{off}),
the rotation angles $g_L$ and $g_R$ can be absorbed into the fermions
$\eta$ via a chiral rotation $\eta_{L,R} \rightarrow g_{L,R}
\eta_{L,R}$. This produces a model of $k$ $SU(2)$ fermions coupled to
a background gauge field $A_+=g_L^{-1}\d_+ g_L$, $A_-= g_R^{-1}\d_-
g_R$.  Via the standard chiral anomaly argument, that can be trusted
for small $r = e^{\varphi}$, 
one derives that the one-loop effective action includes
an $SU(2)$ WZW-model for the field $g$ with Kac-Moody level given by the
five-brane number $k$. We see in particular that the non-zero
background three-form field $H$ in (\ref{H3}) is reproduced by the one-loop
determinants as the Wess-Zumino term of this action.

Thus we seem to arrive quite naturally at the CFT description  
of the type IIA string moving close to the fivebrane, as
derived in \cite{callanetal} using more standard methods. This
action becomes exact for $e^\varphi \ra 0$ and consists of a level 
$k$ $SU(2)$ WZW model combined with a Feigin-Fuchs scalar field $\varphi$
\be
S(x^I) = S_{WZW}(g) + S_{FF}(\varphi). 
\ee
The background charge of the scalar $\varphi$ represents a 
space-time dilaton field that grows linearly for $\phi \ra - \infty$, 
and is normalized such that the total central charge is $c=6$.

\newsubsection{Nature of the horizon}

What is perhaps the most interesting lesson from the description of
black holes in matrix theory, is that the world-sheet CFT, that is
usually taken as the {\it classical} starting point in perturbative string
theory, now arises as an effective quantum description. 
This emergence of CFT as an effective low-energy
theory a la Seiberg-Witten is suggestive of the picture that was put
forward in \cite{strominger}, where the singularities in the CFT
moduli space were attributed to D-branes degrees of freedom that become 
massless at the conifold points. The above derivation of the black hole
geometry shows that this interpretation of CFT as an effective description
is a generic feature of matrix theory. 

Quite generally, the effective world-sheet theory near any object is
induced by integrating out off-diagonal matrix elements
\cite{li-martinec}.  The interpretation of these off-diagonal elements
is somewhat mysterious in the IIA context, since they represent
non-perturbative degrees of freedom. They are mapped to the
fundamental open strings under the duality with the D-string probe
analysis. The distinguishing feature of black hole configurations is
that these off-diagonal degrees of freedom can become massless (on the
string world-sheet) when the string approaches the black hole
horizon. In fact, the appearance of these extra massless modes can be
used to characterize the location as well as the shape of the geometry
near the horizon. Indeed by including the off-diagonal modes as
fundamental excitations, the horizon geometry resolves into a flat
space of a non-abelian nature.  It is tempting to speculate that all
singular classical and quantum features of black holes can be resolved
via such a mechanism.

\bigskip

{\noindent \sc Acknowledgements}

We acknowledge useful discussions with L. Alvarez-Gaum\'e, S. Das,
M. Dekker, F. Larsen, J. Maldacena, G. Mandal, and S. Wadia. This
research is partly supported by a Pionier Fellowship of NWO, a
Fellowship of the Royal Dutch Academy of Sciences (K.N.A.W.), the
Packard Foundation and the A.P. Sloan Foundation.

\vspace{2cm}
\pagebreak[3]
\renewcommand{\theequation}{A.\arabic{equation}}
\setcounter{equation}{0}
{\noindent\sc Appendix  A: The five-dimensional black hole}
\nopagebreak
\vspace{2mm}
\nopagebreak

In this appendix we collect some facts about 5D black holes that are
used in the main text. The black hole metric of \cite{cvetic,horowitz} is
parametrized by a size $r_0$ and three hyperbolic angles
$\alpha,\gamma,\sigma$. The boost parameter $\sigma$ labels a Lorentz
transformation in the $(x^9,t)$-plane. In this parametrization the
solution takes the form
\ba
ds^2 \is  f_k^{-1/2} f_m^{-1/2}[-dt^2 + dx_9^2 + {r_0^2 \over r^2}(\cosh\sigma\, dt
+ \sinh \sigma\, dx_9)^2 ] \nonu
&& \qquad + f_k^{-1/2} f_m^{1/2} [dx_6^2 + \ldots + dx_8^2] \nonu
&& \qquad \qquad + f_k^{1/2} f_m^{1/2} 
[{dr^2\over 1-r_0^2/r^2} + r^2 d^2 \Omega_3],
\ea
with
\ba
f_k(r) \is (1+ {r_0^2\sinh^2 \alpha \over r^2}),\nonu
f_m(r) \is (1+ {r_0^2\sinh^2 \gamma \over r^2}).
\ea
The mass $M$ and Bekenstein-Hawking geometric entropy $S_{BH}$ of such
a black hole are (written in terms of the five-dimensional Planck
length $\ell_p$)
\ba
M \is {r_0^2 \over \ell_p^3} (\cosh 2\alpha + \cosh 2\gamma + \cosh
2\sigma),\nonu 
S \is 2\pi {r_0^3\over \ell_p^3} (\cosh\alpha\, \cosh\gamma\,
\cosh\sigma).
\ea
The black hole carries three charges that can be expressed as
\ba
q_m \is {r_0^2\over \ell_p^2} \sinh 2\alpha,\nonu
q_k \is {r_0^2\over \ell_p^2} \sinh 2\gamma,\nonu
q_N \is {r_0^2\over \ell_p^2} \sinh 2\sigma.
\ea
Note that the quantization of these charges depend on the moduli
describing the string compactification. We have dimensionless
constants $\lambda_m,\lambda_k,\lambda_N$ such that 
\be
q_m = \lambda_m m,\qquad q_k = \lambda_k k,\qquad q_N = \lambda_N N,  
\ee
with positive integers $m,k,N$. The moduli dependent 
prefactors satisfy the condition
\be
\lambda_N\lambda_m \lambda_k=1.
\ee
The extremal BPS black hole is obtained in the limit $r_0 \ra 0,$
$\alpha,\gamma,\sigma \ra \infty$, while keeping the above charges
fixed. In this limit the mass formula simplifies to
\be
M_{ext} = (\lambda_m m + \lambda_k k + \lambda_N N)/\ell_p,
\label{ext}
\ee
while the entropy becomes moduli independent
\be
S_{BH} = 2\pi \sqrt{Nmk}.
\ee
In fact,  the limiting behaviour of the non-extremal mass becomes clear
if written as
\be
M = (\lambda_m m \coth 2\alpha + \lambda_k k \coth 2\gamma + \lambda_N N
\coth 2\sigma)/\ell_p.
\ee 
The non-extremal entropy can be given a suggestive form by
defining left-moving and right-moving charges
\ba
m_R \is m_L e^{-4\alpha},\nonu
k_R \is k_L e^{-4\gamma},
\nonu N_R \is N_L e^{-4\sigma}
\ea              
with $m_L-m_R=m$ etc. Using this notation the non-extremal mass and
entropy are
\ba
M \is \Bigl( \lambda_N (N_L + N_R) + \lambda_m (m_L + m_R) +
\lambda_k (k_L + k_R)\Bigr)/\ell_p
\label{mass} \\[3mm]
S_{BH} \is 2\pi
\left(\sqrt{N_L} + \sqrt{N_R}\right)
\left(\sqrt{m_L} + \sqrt{m_R}\right)
\left(\sqrt{k_L} + \sqrt{k_R}\right)
\label{ent}
\ea
In the extremal limit $k_R \ra 0,$ $k_L \ra k$ etc.

In the setup of this paper, the black hole arises in the type IIA
theory with string coupling $g_s$ compactified on a circle of radius
$R_9$ times a four-torus with volume $V$. In terms of these moduli the
5-dimensional Planck length becomes
\be
\ell_p=(R_9V/g_s^2\alpha'{}^4)^{1/3}.
\ee 
The prefactors $\lambda_m,\lambda_k,\lambda_N$ that measure the charge
quantization of what is now interpreted as string winding number,
fivebrane wrapping number and momentum, can be found by comparing the
duality invariant mass formula (\ref{ext}) with the expression
(\ref{extmass}).

One obtains 6-dimensional black string solutions by decompactifying
the $x_9$ direction. In the absence of gravitational waves that can
propagate along the string \cite{gary}, these solutions are
characterized by only three parameters. The boost parameter $\sigma$,
and the corresponding distinction between the left- and right-momentum
$N_L$ and $N_R$ disappears, because the Lorentz boost along the string
direction becomes a symmetry of the solution.

\vspace{1cm}
\pagebreak[3]
\renewcommand{\theequation}{B.\arabic{equation}}
\setcounter{equation}{0}
{\noindent\sc Appendix B: Supersymmetry and perturbative BPS spectrum}
\nopagebreak
\vspace{2mm}
\nopagebreak

In this appendix we describe the the realization of the supersymmetry
and the counting of perturbative BPS states in the $S^1$
compactification of the type IIA string to nine dimensions in terms of
the finite $N$ matrix string.

In the matrix string the $D=10$ supersymmetry algebra is realized
along the following lines. The broken supercharges $\tilde Q$ are
given by the zero-modes of the world-sheet fermion fields $\theta$,
while the unbroken supercharges $Q$ are realised by the world-sheet
${\cal N}=8$ supersymmetry generators $G$. The space-time algebra is
then recovered by implementing the appropriate level-matching constraint 
and mass-shell condition. 

It is convenient to use a chiral $SO(8)$ notation, with $\a,\adot$ denoting 
the ${\bf 8}_s$ and ${\bf 8}_c$ representation respectively. 
Concretely, the expessions for the supersymmetry charges in terms of 
the world-sheet fields are
\ba
\label{susy}
\tilde{Q}^\a \is \sqrt{{p^+\LL \over N}} 
\oint  \!d\sigma\, \tr \,\th^\a,\nonu
Q^\adot \is \sqrt{N\over{p^+\LL} }
\oint  \!d\sigma\,\tr\Bigl\lbrack \theta^T \left(E 
+\gamma^i ( \Pi_i \!+ \!DX_i) 
+ \gamma^{ij}[X_i,X_j]\right)\Bigr\rbrack^\adot. 
\ea
Similar expressions hold
for the right-movers with $p^L_\pm = p_0 \pm p_9^L$ 
replaced by $p^R_\pm = p_0 \pm p_9^R$, where $p_9^{L,R}=p_9 \pm w_9$. 
%\ba
%\tilde{Q}^\adot \is \oint  \!d\sigma\, \tr \tilde\th^\adot,\nonu
%Q^\a \is \oint  \!d\sigma\,\tr\Bigl(\tilde\theta^T (E 
%\gamma^i DX_i - \gamma^i\Pi_i - \gamma^{ij}[X_i,X_j])\Bigr)^\adot 
%\ea
The generators $Q^\adot$ satisfy the algebra
\ba
\label{salgebra}
\{\tilde Q^\a, \tilde Q^\b\} \is p{}\LL^+ \delta^{\a\b}, \nonu
\{Q^\a,\tilde Q^\bdot\} \is 0, \nonu
%\{Q^\adot,Q^\bdot\} \is p{}\LL^-
%\delta^{\adot\bdot}
%\ea 
%\be
\{Q^\adot, Q^\bdot\} \is {N\over p^+_L}(H+P) \delta^{\adot\bdot},
\ea
with $H$ and $P$ the total world-sheet energy and momentum, 
as defined in (\ref{H}) and (\ref{P}). The algebra (\ref{salgebra})
becomes identified with the space-time supersymmetry algebra
provided we impose the level-matching and mass-shell relations
\ba 
N (H + P) \is p\LL^+ p\LL^-, \nonu N(H - P) \is p\RR^+ p\RR^-, 
\ea
or equivalently
\ba
NH \is p_0^2 - p_9^2 - w_9^2,\nonu
NP \is p_9w_9.
\ea

The BPS spectrum of perturbative type II string states arises as
follows.  In this case the spinors $\epsilon,\tilde\epsilon$
satisfy the chirality conditions
\be
\gamma_9\epsilon=\pm\epsilon,\qquad \tilde\epsilon=0.
\ee
where the sign is determined by that of $p_9w_9$.
The BPS equation of motion becomes in this case
\be
(1-\gamma^9)(\gamma^9 E 
+\gamma^{9i} DX_i + \gamma^i\Pi_i + \gamma^{ij}[X_i,X_j]) = 0
\ee
which is decomposed as
\be
P^i + DX^i=0, \qquad [X^i,X^j]=0, \qquad E= 0.
\ee
The last two equations imply that BPS states are necessarily free
string states, while the first equation tells us that
only left-moving fields are allowed. 

Since we consider only left-moving excitations, we have $\Lbar_0=0$
which implies $p_-^R=0$. This in turn gives the relation $p_+^L=p_9$
showing that for BPS states, one can identify the (left-moving)
light-cone momentum with the pure spatial momentum $p_9$.  Similarly
we have $p_-^L=w_9$.  Level-matching now simply relates the $L_0$
eigenvalue to the total winding number as $H+P = p_9w_9$. Similarly,
for the individual strings we have $L_0^{(i)} = p^{(i)}_9 w^{(i)}_9$.

The spectrum of these perturbative BPS string states can be computed
exactly from the chiral partition function $\chi(\R^8;q)=\Tr \,
q^{L_0}$ of the superstring
\be
\chi(\R^8;q) = \sum_n d(n) q^n= \prod_n \left(1+q^n\over 1-q^n\right)^8.
\ee
by using the general result described in \cite{orbifold}. One obtains
\be
\sum_N p^N \chi(S^N\R^8;q)= \prod_{n,m}(1-p^nq^m)^{-d(nm)}.
\ee
In terms of perturbative type II string theory this describes a gas of
perturbative BPS states. Note that (only) for compactifications on
$S^1$ such a superposition of single-string BPS states is again BPS,
since the right-moving compactified momenta always point in the same
direction.

\renewcommand{\Large}{\large}

\end{document}